\documentclass{egpubl}
\usepackage{eurovis2023}

\SpecialIssuePaper         

\CGFStandardLicense

\usepackage[T1]{fontenc}
\usepackage{dfadobe}  

\usepackage{cite}  
\BibtexOrBiblatex
\electronicVersion
\PrintedOrElectronic
\ifpdf \usepackage[pdftex]{graphicx} \pdfcompresslevel=9
\else \usepackage[dvips]{graphicx} \fi

\usepackage{egweblnk}

\usepackage{algorithm}
\usepackage{algpseudocode}
\usepackage{amssymb}
\usepackage{relsize}
\usepackage{booktabs}
\usepackage{tabu}
\usepackage{multirow}
\usepackage[capitalise]{cleveref}
\usepackage{listings}
\usepackage{xcolor}
\usepackage{stackengine}
\usepackage{tikz}
\usepackage{balance}

\newif\ifsubmission

\definecolor{codegreen}{rgb}{0,0.6,0}
\definecolor{codegray}{rgb}{0.5,0.5,0.5}
\definecolor{codepurple}{rgb}{0.58,0,0.82}
\definecolor{backcolor}{rgb}{0.95,0.95,0.92}

\usepackage{adjustbox}
\usepackage{xspace}
\usepackage{relsize}
\def\code#1{{\mbox{\relsize{-1}\texttt{#1}}}\xspace}

\crefname{lstlisting}{Listing}{Listings}

\usepackage{caption}
\lstdefinestyle{overleaf}{
    backgroundcolor=\color{backcolor},
    commentstyle=\color{codegreen},
    keywordstyle=\color{magenta},
    numberstyle=\tiny\color{codegray},
    stringstyle=\color{codepurple},
    basicstyle=\ttfamily\footnotesize,
    breakatwhitespace=false,
    breaklines=true,
    captionpos=b,
    keepspaces=true,
    numbers=left,
    numbersep=5pt,
    showspaces=false,
    showstringspaces=false,
    showtabs=false,
    tabsize=2
}
\title[Immersive ExaBrick]{Immersive ExaBrick: Visualizing Large AMR Data in the CAVE}
\ifsubmission
\author[Submission ID XXXX]{Submission ID: XXXX}
\else
\author[Z.~Wang \& S.~Wesner \& S.~Zellmann]
{\parbox{\textwidth}{\centering Zhaoyang~Wang$^{1}$\orcid{0009-0007-8606-6826}
    Stefan~Wesner$^{1}$\orcid{0000-0002-7270-7959},
    Stefan~Zellmann$^{1}$\orcid{0000-0003-2880-9090}
        }
        \\
{\parbox{\textwidth}{\vspace{-2em}\centering $^1$University of Cologne
       }
}
\vspace{-0.5em}
}
\fi

\newif\ifdiff

\usepackage{soul}
\ifdiff
\definecolor{midgreen}{rgb}{0,0.6,0}

\def\removed#1{\textrm{\color{red}\st{#1}}}
\else

\def\removed#1{}
\fi

\usepackage{graphicx}
\usepackage{subcaption}
\usepackage{mwe}

\teaser{
 \centering
 \includegraphics[width=18cm]{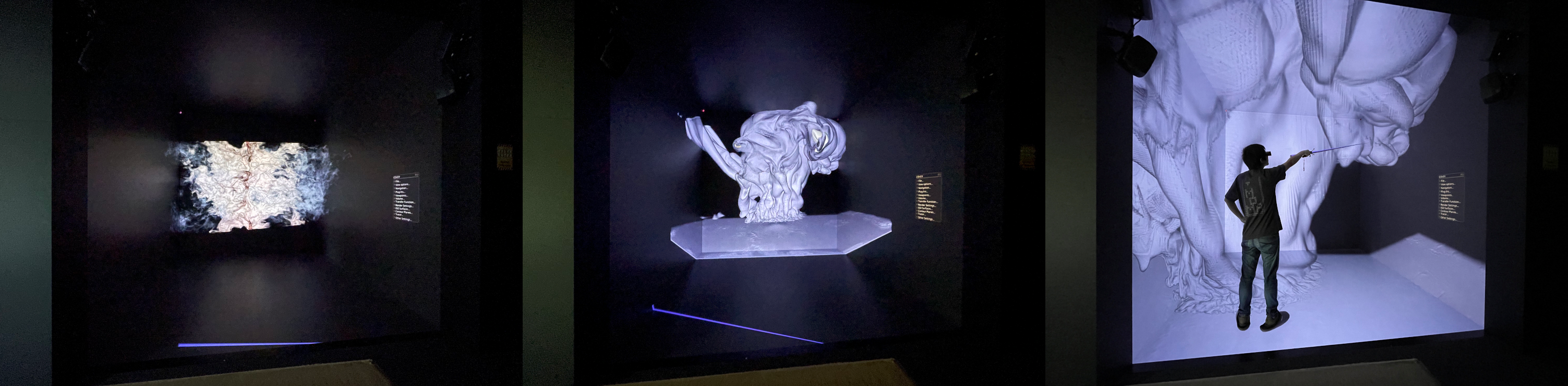}
\vspace{-1.0em}
 \caption{\label{fig:teaser}%
 Real-time rendering of AMR data sets with our tool, in the CAVE virtual environment at the University of Cologne. Left: molecular cloud data set; Mid: Meteor impact data set (t=46,122s); Right: Meteor impact zoomed in.}
}

\begin{document}

\maketitle
\begin{abstract}
Rendering large adaptive mesh refinement (AMR) data in real-time in virtual reality (VR) environments is a complex challenge that demands sophisticated techniques and tools. The proposed solution harnesses the ExaBrick framework and integrates it as a plugin in COVISE, a robust visualization system equipped with the VR-centric OpenCOVER render module. This setup enables direct navigation and interaction within the rendered volume in a VR environment. The user interface incorporates rendering options and functions, ensuring a smooth and interactive experience. We show that high-quality volume rendering of AMR data in VR environments at interactive rates is possible using GPUs.
\end{abstract}

\section{Introduction} \label{sec:intro}
In recent years, a trend can be observed for scientific simulations to generate volumetric data in the form of adaptive mesh refinement (AMR) topologies.
AMR data~\cite{berger:1984:adaptive,berger:1989:local} is characterized by its hierarchical structure, where the computational domain is divided into a series of hierarchical grids with varying levels of refinement. It offers the advantage of enhanced simulations by enabling higher resolution in critical regions while maintaining a coarser representation in less significant regions.
With the help of the ExaBrick framework~\cite{wald2021exabrick}, large AMR data can be re-arranged into the ExaBricks data structure. This restructuring of data enables scientific visualization
of such data at interactive framerates.
However, real-time visualization of large AMR data sets presents formidable challenges due to their intricate nature and the imperative for interactive exploration and analysis.

Real-time rendering technology assumes a pivotal role in enabling interactive exploration and analysis
in virtual reality (VR) environments,
allowing users to seamlessly navigate and manipulate data in a responsive manner. VR has become a powerful technology that provides an immersive and interactive experience, making it an ideal platform for visualizing complex scientific data. By harnessing the potential of VR, researchers and scientists can attain a profound comprehension of the complex structures and phenomena represented by AMR data. 

To create an immersive experience that seamlessly integrates users' natural head movements, enables exploratory interactions, and facilitates the manipulation of data sets within the VR environment,
rendering frameworks are required to implement virtual stereo cameras to generate
viewports from; in contrast to camera models used for non-immersive applications,
VR camera models support head tracking by placing the camera position at an off-center
position.
These off-axis cameras allow for the camera to be positioned at an angle that does not rigidly align with the center of the scene or object. Leveraging the advanced capabilities of the CAVE system~\cite{cruz-neira} installed at the University of Cologne and the VR renderer OpenCOVER, which is integrated within the COVISE platform, we can expand upon the ExaBrick framework. Our extensions allow us to harness the power of 3D rendering within VR environments, promising a groundbreaking leap in immersive data visualization and interaction.

Our main contributions are:
\begin{itemize}
    \item A real-time rendering solution for large AMR data sets in a VR environment;
    \item An analysis of the lessons learned and challenges to integrate the ExaBrick framework~\cite{wald2021exabrick} as a plugin into COVISE;
    \item An immersive VR application that enables users to navigate and interact with rendered data using head trackers or controllers, and access and manipulate rendering options.
\end{itemize}

\section{Related Work} \label{sec:related}

Techniques to visualize large-scale data become paramount, and there has been a lot of development in this area~\cite{sarton:2023:state}. For example, Sarton et al.~\cite{sarton:20:gpu-vis} introduced an efficient solution for managing large 3D grid data on a GPU. This approach minimizes memory transfers between GPU and host and is designed to maintain GPU performance while reducing CPU-GPU communication. Morrical et al.~\cite{morrical:20:bilinear-elements}  harnessed ray tracing hardware in NVIDIA's RTX GPUs, achieving substantial performance enhancements in unstructured mesh point location. 

Rendering AMR data has been a subject of significant interest and importance in the scientific visualization community. Berger et al.\ brought up with the AMR data structure~\cite{berger:1984:adaptive,berger:1989:local}, which enables simulations to focus on more interesting parts in the data. The cells the data is composed of have different refinement levels according to their relative importance. The more important the area is, the smaller are the cells in this area. AMR data comes in different forms, such as block-structured AMR~\cite{colella:2009:chombo} or Octrees~\cite{burstedde:2011:p4est}.

Kähler et al.~\cite{kahler2003interactive,kahler2006gpu} proposed to render AMR data on the GPU via a data structure that discards the original AMR hierarchy and builds blocks containing only same-refinement-level cells. While their works excel in supporting nearest neighbor reconstruction and vertex-centered AMR, challenges of smooth interpolation for cell-centered data resulting from T-junctions remain unaddressed. These challenges were addressed by Weber~\cite{weber2012efficient} on the CPU by generating stitch cells in parallel using extra layers of cells around boundaries. Wald et al.~\cite{wald2017cpu} proposed to use tent-shaped basis function interpolation, enabling smooth interpolation including at refinement level boundaries. Wang et al.~\cite{wang:18:iso-amr} presented reconstruction filters utilizing the octants of dual AMR cells and operate directly on cell-centered AMR data. 

These prior works by Wald and Wang primarily focused on CPU reconstruction. Their contributions involved filters that demanded expensive cell location calculations through KD-trees, requiring eight such look-ups per ray marching step to interpolate the samples. This process becomes exceptionally costly when executed on GPUs due to their limited caching capabilities and their high concurrency, which can lead to congestion on the memory subsystem.

To address these issues, Wald et al.~\cite{wald2021exabrick} later introduced a significant advancement in this field through the introduction of the ExaBrick data structure and rendering framework. The paper delves into an in-depth exploration of practical models and rendering techniques tailored for AMR data sets.
A detailed description is provided in \cref{sec:background}.
The focal point is a software solution called \emph{ExaBrick}, which will also form the basis for our work.
The proposed framework extends earlier work by Kähler and others~\cite{kahler2003interactive,kahler2006gpu}
but lifts the restriction that their GPU data structure is only applicable to nearest neighbor interpolation or vertex-centric
data. Another notable property is ExaBrick's use of hardware ray tracing cores via OptiX~\cite{parker2010optix}.


Zellmann et al.'s~\cite{zellmann:2022:cise} subsequent extension adapts ExaBrick to support steady flow visualization using particle tracing. Later, Zellmann et al.~\cite{zellmann2022design} harnessed modern workstations and APIs to visualize exa-scale time-varying AMR data sets and achieved smooth animation at interactive rates. Addressing the challenges coming with scattering events and global illumination, Zellmann et al.~\cite{zellmann-beyond-exabricks} implemented volumetric path tracing for AMR data using Woodcock tracking~\cite{woodcock1965techniques}, where ExaBrick serves as an acceleration data structure supplying density majorants.

Various visualization techniques have been explored in the context of VR environments, aiming to enhance user experiences and enable immersive interactions with complex data and simulations ~\cite{sherman03}. In seminal work from 1998, Cruz-Neira et al.~\cite{cruz-neira} introduced the CAVE virtual reality and scientific visualization system. A descendent of such systems, yet realized with modern technology such as active stereo backprojection at high resolution, or high-precision tracking with infrared markers, was recently installed at the University of Cologne~\cite{cave-cologne}.
CAVE systems use off-axis perspective projection techniques, which are required to implement 6-DoF stereo camera systems with arbitrarily positioned viewing centers.

Horan et al.~\cite{horan:2018:cave-like} for example introduced a versatile CAVE-like VR system. This system integrates a 6-DoF haptic interaction system and immersive 3D surround sound audio, offering a highly immersive experience. A standout feature is the INCA 6D haptic system developed by Haption GmbH~\cite{perret:09:inca6d}, enabling users to grasp and feel objects with up to 37.5N force feedback on a gripping tool. The paper discusses the system's pros and cons across various configurations and VR applications.

Head-mounted displays (HMDs) are an alternative to CAVE-like VR and are used widely in this field. For example, the Oculus Rift headset and Oculus Touch controllers are used in the study by Kalarat and Koomhin~\cite{kalarat2019real} with a VR application for real-time volume rendering interaction with 1D transfer functions, enabling users to visualize stereoscopic images at 60 frames per second.

Utilizing the advantages of VR techniques, scientists are able to visualize data in different domains; Koger et al.~\cite{koger2022virtual}, e.g., investigated the potential of VR as an immersive platform for interactive medical analysis, enabling the manipulation and exploration of CT images of the cardiopulmonary system in a 3D environment, facilitating the generation of new data analysis perspectives and enhancing data interpretation for medical practices including training, education, and investigation. Recently, Nam et al.~\cite{ospray-immersive} introduced an extension to OSPRay Studio, enabling 3D virtual environment display on tiled walls. This extension employs gesture-based interaction, utilizing Microsoft Kinect sensors to track user movements and update 3D objects based on interpreted gestures.

The COVISE visualization framework~\cite{rantzau1996collaborative}, along with its COVER module~\cite{rantzau1998covise} (later OpenCOVER), serves as a pivotal avenue for rendering scientific data within VR environments. This integration of virtual reality with scientific visualization leverages OpenSceneGraph~\cite{osg} and OpenCOVER, enabling real-time 3D rendering in COVISE. COVISE's versatility is highlighted by the incorporation of various rendering approaches, exemplified by Schulze et al.'s volume plugin~\cite{schulze2001volume} and by Zellmann et al.'s Visionaray library~\cite{visionaray}, which, notably, inspired the development of our ExaBrick plugin.


\section{Background: ExaBrick Data Structure and Framework}
\label{sec:background}
Our contribution builds considerably upon ExaBrick by Wald et~al.~\cite{wald2021exabrick}. ExaBrick can be considered a data structure over
the basis function method presented in earlier work by Wald et al.~\cite{wald2017cpu},
as well as a framework based on NVIDIA OptiX that makes use of hardware ray tracing.
The fundamental problems solved are those of smooth interpolation including at the
level boundaries---this is achieved by using tent basis functions---as well as coherent
traversal with viewing rays, in the spirit of the data structure proposed by Kähler
et al.~\cite{kahler2006gpu}.

Other than Kähler's data structure, the integration domains chosen by Wald do overlap
(as they include one half cell's overlap per side to accommodate the tent basis); Wald et al.\
address this issue by computing a domain decomposition over the regions where the
basis function domains overlap, so a volume renderer performing absorption and emission
integration is still presented with non-overlapping integration regions.
Each domain of the domain decomposition stores pointers
to a set of ``bricks'' (\emph{these} are the same as Kähler's, and span the actual \emph{data}
domain of the cells). By interactively classifying and culling empty regions, only those
regions that are visible become active, hence the term \emph{active brick regions} (ABR)
for the axis-aligned boxes constituting the domain decomposition. The ABRs can be traversed
similarly to the original bricks by Kähler, allow for coherent accesses amenable to GPUs,
and serve as an adjacency data structure required for smooth level boundaries.

ExaBrick is however also an open source framework\footnote[1]{\url{https://github.com/owl-project/owlExaBrick}}
organized into what can semantically considered to
be a software library (depending on OptiX, and on OWL, the OptiX Wrappers Library~\cite{wald2022owl}),
and a viewer application (\code{exaViewer}) that uses the library. Technically, the library and
viewer are statically compiled into the same executable, but C++ classes define the interface
and allow for easy separation of the two components, so that the library part could, e.g., be
compiled into a Unix shared object or a Windows dynamic link library. The main interface of the
``library portion'' is the \mbox{\code{OptiXRenderer}} class, which also wraps the virtual camera,
consumes and populates a (CUDA device) pointer to the frame buffer, and has pointers to the data
that is supplied via a proprietary (yet documented and simple) file format and config file
facilities.

\section{Method Overview} \label{sec:overview}
We propose to use ExaBrick to drive immersive rendering in the CAVE by building
it as a shared library and integrating it into OpenCOVER, COVISE's rendering
component, as a \emph{viewport plugin}. Viewport plugins capture the renderer's
tracking state, transform that into camera parameters that can be consumed and
finally turned into rendered images using a custom rendering method, to
eventually be displayed in OpenCOVER's viewport.

The \code{exaViewer} application developed by Wald et al.\ forms the viewing tool and GUI of ExaBrick.
For VR, however, this tool is only of limited use.
It relies on an on-axis pinhole camera placed in direct alignment with the scene's center. Consequently, it captures scenes from a viewpoint where the camera's optical axis is congruent with the line of sight to the center of the scene. This yields a straightforward and centrally-focused perspective of the scene, but only for desktop applications with mono viewing and without headtracking interaction.

The original camera needs to be replaced with an off-axis model to support CAVE-like stereo. In ray tracing we use rays to simulate the line of sight from the eyes to the object to simulate light transport and light's propagation to the human eye bouncing off objects. The new camera in the plugin
should generate primary rays following the off-axis camera model.



\section{Implementation} \label{sec:impl}
The main programming language used for the implementation is C++. Important code snippets will be presented here, where applicable in C++, or in an abridged form of that where clarity demands this; complete code is available on a development branch on GitHub
\footnote[2]{\url{https://github.com/zywang3/covise/tree/zhaoyang}}.

\subsection{OpenCOVER Viewport Plugin Control Flow}
OpenCOVER can be extended using plugins, which are dynamic libraries implementing a function interface that OpenCOVER calls at runtime. To create a plugin, we inherit from the C++ class \code{coVRPlugin} and implement its virtual methods. Some of the virtual methods, when called,
guarantee that an OpenGL context is ``current'', i.e., when issued from there, OpenGL calls by our implementation are executed on the
main graphics thread.

The interface to obtaining tracking parameters for the user's position as mentioned in \cref{sec:overview} is supported by accessing OpenCOVER's OpenGL state. This is realized using the C++ utility class \code{MultiChannelDrawer}; the multi-channel drawer class also allows the plugin to write RGBA and depth pixels to each \emph{channel}; channels form framebuffer abstractions, e.g., for the left/right eye, per projection wall, of the CAVE environment.
This workflow of accessing OpenCOVER's camera transforms and directly writing to the frame and
depth buffer allows us to realize what is called a \emph{viewport plugin}---which is opposed
 to
 other plugin types that, e.g., extend the scene graph structure of the VR renderer.

\begin{figure}[htpb]
  \centering
  \includegraphics[width=7.8cm]{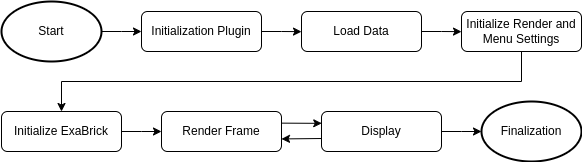}
  \caption{Flowchart depicting the structure of our plugin.}
  \label{fig:Flowchart_Plugin}
\end{figure}

The ExaBrick plugin follows a structured workflow that includes Plugin initialization, data loading, setting up initial values and menus, and initialization of the ExaBrick library.
OpenCOVER then enters the render loop, giving the plugin the chance to render frames and display them. On shutdown, the plugin executes
a finalization phase to tear down any state, including that of the ExaBrick library. See \cref{fig:Flowchart_Plugin} for an overview.

\subsection{ExaBrick Application/Plugin--Library Interface}
The ExaBrick plugin utilizes various functions and tools from the original ExaBrick library. We build the ExaBrick tools and \code{exaViewer} executable locally by integrating the whole C++ project as a \code{git} submodule into COVISE.
We build the rendering logic into a shared library;
interoperability between COVISE/OpenCOVER and ExaBrick is ensured
as both software projects use \code{CMake} as the build system.

ExaBrick imports its data via a set of proprietary, yet documented, file formats.
The corresponding code for loading data is encapsulated in the \code{loadFile} function using COVISE's \code{FileHandler} class. The file handler of our plugin is defined to read text-based config files, which
point to binary input files with scalar data and bricks; as described in the original
publication~\cite{wald2021exabrick}, any
cartesian AMR format with compatible data can be converted to this format using tools coming
with the ExaBrick framework.
When the file handler gets invoked,
loading these files is realized using ExaBrick functions that we promoted to API functions
that can be called at the library/application interface.

During the \emph{Initialize Render and Menu Settings} phase (cf.\ \cref{fig:Flowchart_Plugin}), a multitude of parameters for the renderer are initialized. This encompasses setting up transfer function values, configuring rendering preferences, and determining the ideal ray marching step size $dt$.
A detailed description of the menu structure is found in \cref{sec:menus}.
These settings are established directly after the creation of \code{OptiXRenderer}, which is the central class of the ExaBrick library.

Afterwards we create the \code{MultiChannelDrawer}, the API that allows us to realize viewport plugins and render into COVER's viewport. The class can handle various rendering modes, update geometry, and handle the data for each view independently or collectively, depending on the selected view settings. The camera matrices are obtained from the \code{MultiChannelDrawer} (Line~2--6 in \cref{lst:mcd})  and passed to the renderer (Line~14--22 in \cref{lst:mcd}) using the function \code{setCameraMat()}. Furthermore, we obtain pointers the depth and color buffer using the \code{depth()} and \code{rgba()} member functions of the \code{MultiChannelDrawer} class (Line~26--29 in \cref{lst:mcd})
that our renderer can write to directly. After rendering a single frame (\cref{lst:mcd}, Line~32),
the final phase responsible for displaying the resulting images, now resident in the
depth and color buffers, is initiated by calling the \code{MultiChannelDrawer}'s
\code{swapFrame} function (\cref{lst:mcd}, Line~35).

\begin{lstlisting}[language=C++, caption={Interfacing between COVER and our library using the \code{MultiChannelDrawer} C++ class.},label={lst:mcd}]
// Get osg matrices from drawer
osg::Matrix mv
    = multiChannelDrawer->modelMatrix(chan)
    * multiChannelDrawer->viewMatrix(chan);
osg::Matrix pr
    = multiChannelDrawer->projectionMatrix(chan);

// cast osg matrices into ExaBrick-compatible
// types, also switching from row to col-major
math::mat4f view = osg_cast(mv);
math::mat4f proj = osg_cast(pr);

// update camera when matrices changed
if (notsame(view, oldView)
 || notsame(proj, oldProj) {
  // Compute inverse matrices and upload to
  // the device
  plugin->renderer->setCameraMat(
    view, proj);
  oldView = view;
  oldProj = proj;
}

// Pass pointers to frame buffer and depth buffer
// along to the renderer
plugin->renderer->dbPointer
    =  multiChannelDrawer->depth(chan);
plugin->renderer->fbPointer
    =  multiChannelDrawer->rgba(chan);

// Render the frame
plugin->renderer->renderFrame();

// Display the rendered frame
multiChannelDrawer->swapFrame();
\end{lstlisting}

% ...

\subsection{Adjustments to Existing Shader Code}

The original shader code of ExaBrick creates an on-axis camera with the device function \code{generateRay}.
However, since our application requires an off-axis camera, we have extended the functionality of this \code{generateRay} function to calculate rays suitable for our application's needs. 

\begin{figure}[htpb]
  \centering
  \includegraphics[width=9cm]{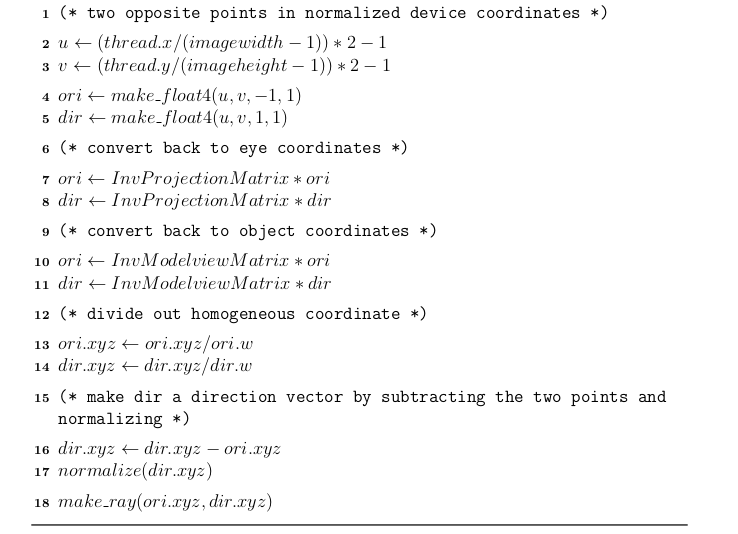}
  \caption{Algorithm calculating camera rays backwards using inverse projection and model-view matrices~\cite{zellmann:phd}.}
  \label{fig:algo-inv}
\end{figure}

As described for example in Zellmann's dissertation~\cite{zellmann:phd}, we generate 
orthogonal rays in normalized device coordinates (NDC) through each pixel, originating at $z=-1$ and
pointing in the positive $z$ direction. By applying the inverse viewing transform (projection and model-view matrices obtained via \code{setCameraMat} in \cref{lst:mcd}), these orthogonal
rays are transformed to world (in our case: CAVE) space and match the camera settings chosen through OpenGL, i.e., if the
chosen OpenGL camera frame exhibits an off-axis frustum, so will the convex hull of the set of camera rays.
The procedure is summarized in \cref{fig:algo-inv} (reprinted with permission from~\cite{zellmann:phd}).


\subsection{VR User Interface} \label{sec:menus}

\begin{figure}[htpb]
  \centering
  \includegraphics[width=8cm]{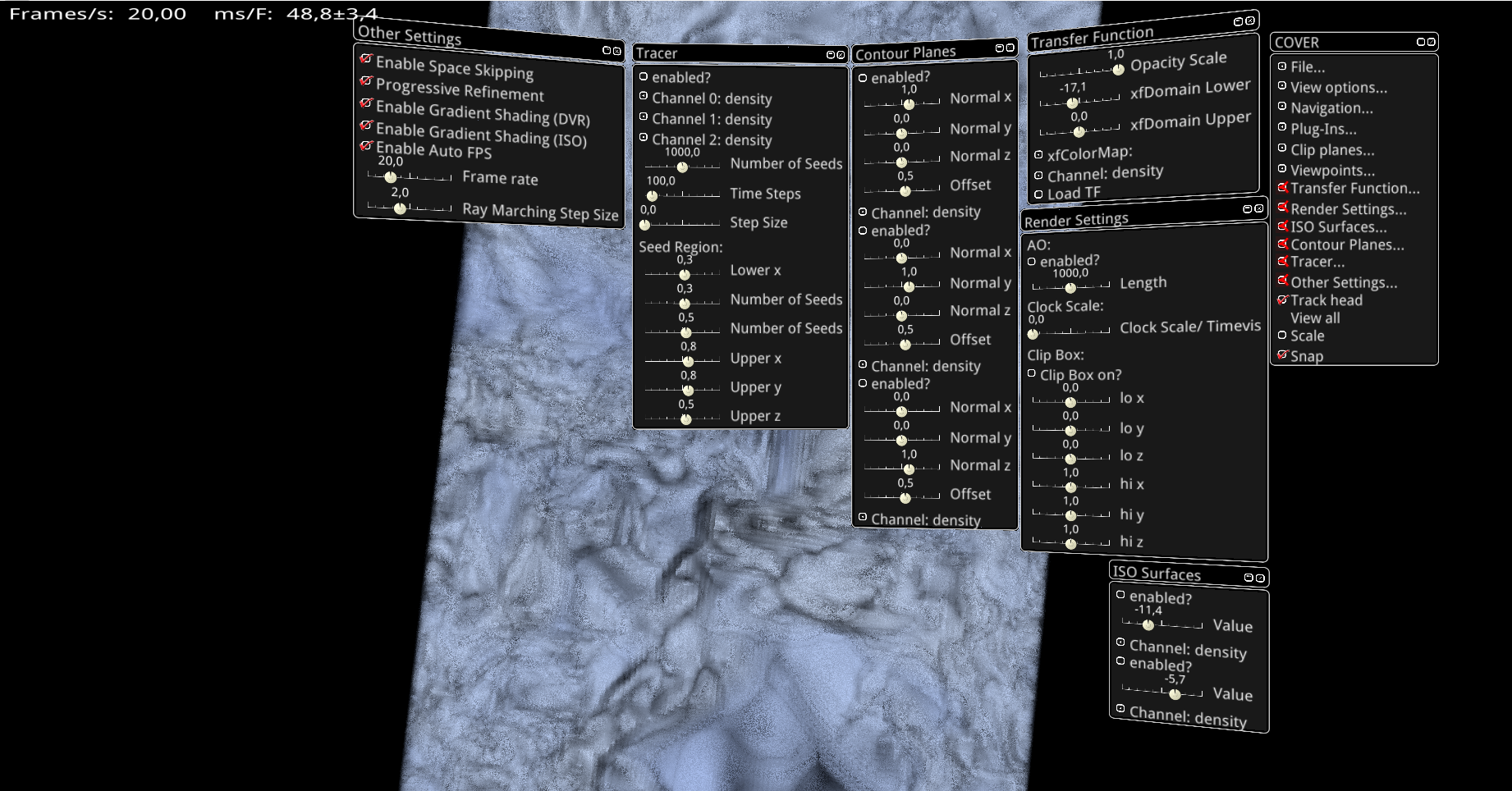}
  \caption{Screenshot of the menu structure of the VR plugin.}
  \label{fig:Menu_Plugin}
\end{figure}

To allow for user interaction with the AMR data in the CAVE,
we expanded the functionality of the VR menu within the OpenCOVER interface by incorporating comprehensive menus and submenus.
These menus largely mirror the menu structure and user interface of the
\code{exaViewer} program.
Typical tasks exposed through the user interface include setting opacity scale, fine-tuning of iso-surface and contour plane parameters, and toggling rendering options.
An overview of the menu structure is given in \cref{fig:Menu_Plugin}.
In the CAVE environment, these menus are manipulated using the
``Flystick~2'' input device.

\section{Results} \label{sec:results}
We performed a quantitative evaluation to assess what quality settings can be
achieved while still retaining interactive framerates for VR.
For the evaluation we use the following hardware and software setup:
five-sided CAVE, $3m^3$ in size, using back projection active stereo (1600 $\times$ 1600
pixels per left and right eye,  each frame is delivered by a separate GPU).
An ART infrared tracker is used for the head position and pointing device (``Flystick~2'').
The system is powered by ten NVIDIA Quadro~RTX~6000 GPU (Turing generation) with 24 GB GDDR memory, as well as Intel Xeon Gold 6130 CPUs @ 2.10GHz, with 128~GB RAM each. The software environment used is based on Ubuntu Linux 20.04, NVIDIA driver version 525.125.06, CUDA 12.0, and NVIDIA OptiX 7.0.

    \begin{figure*}
        \centering
        \begin{subfigure}[b]{0.475\textwidth}
            \centering
            \includegraphics[width=\textwidth]{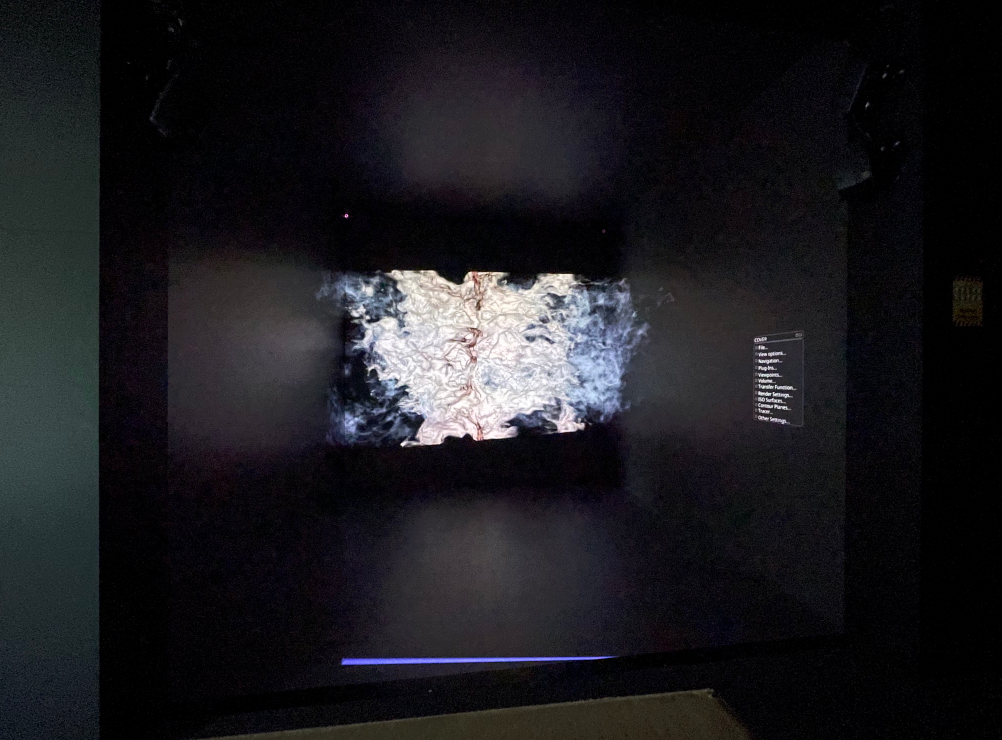}
            \caption[subfig1]%
            {{\small Molecular Cloud viewpoint~1 (zoomed-out)}}    
            \label{fig:molecular_cloud_viewpoint_1}
        \end{subfigure}
        \hfill
        \begin{subfigure}[b]{0.475\textwidth}  
            \centering 
            \includegraphics[width=\textwidth]{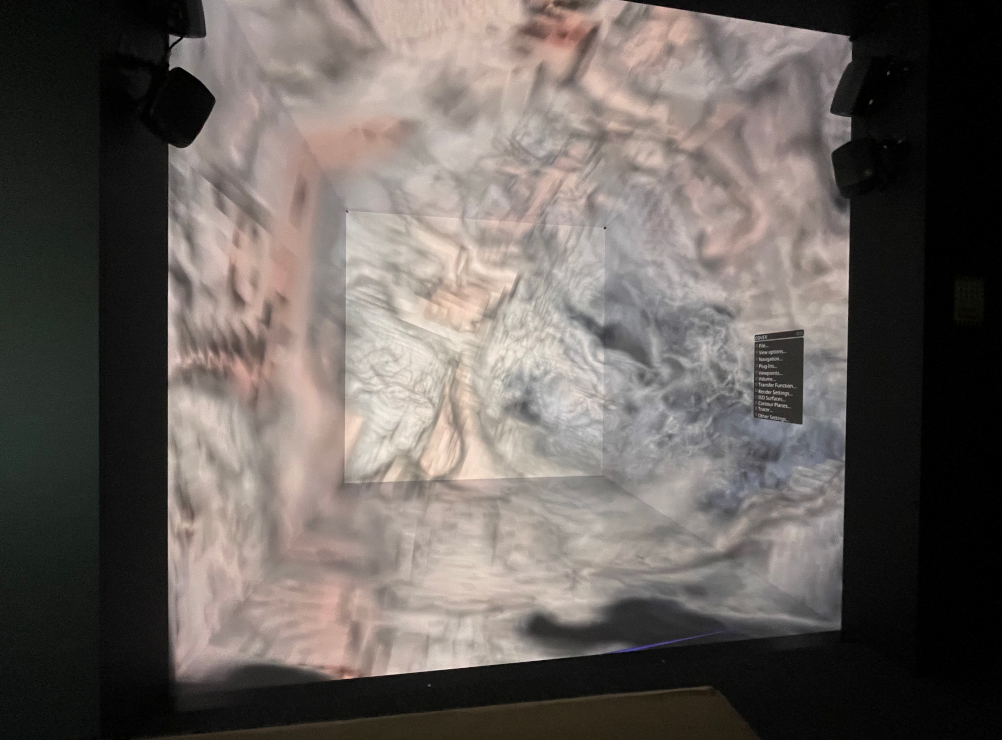}
            \caption[]%
            {{\small Molecular Cloud viewpoint~2 (zoomed-in)}}    
            \label{fig:molecular_cloud_viewpoint_2}
        \end{subfigure}
        \vskip\baselineskip
        \begin{subfigure}[b]{0.475\textwidth}   
            \centering 
            \includegraphics[width=\textwidth]{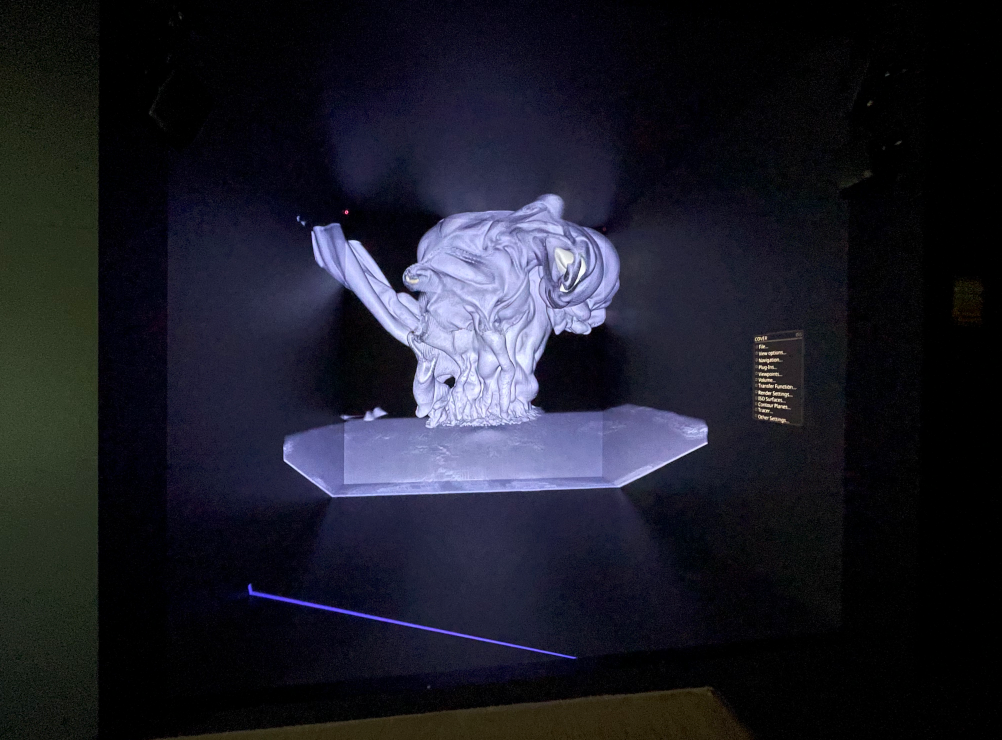}
            \caption[]%
            {{\small LANL Meteor Impact viewpoint~1 (zoomed-out)}}    
            \label{fig:LANL_meteor_impact_viewpoint_1}
        \end{subfigure}
        \hfill
        \begin{subfigure}[b]{0.475\textwidth}   
            \centering 
            \includegraphics[width=\textwidth]{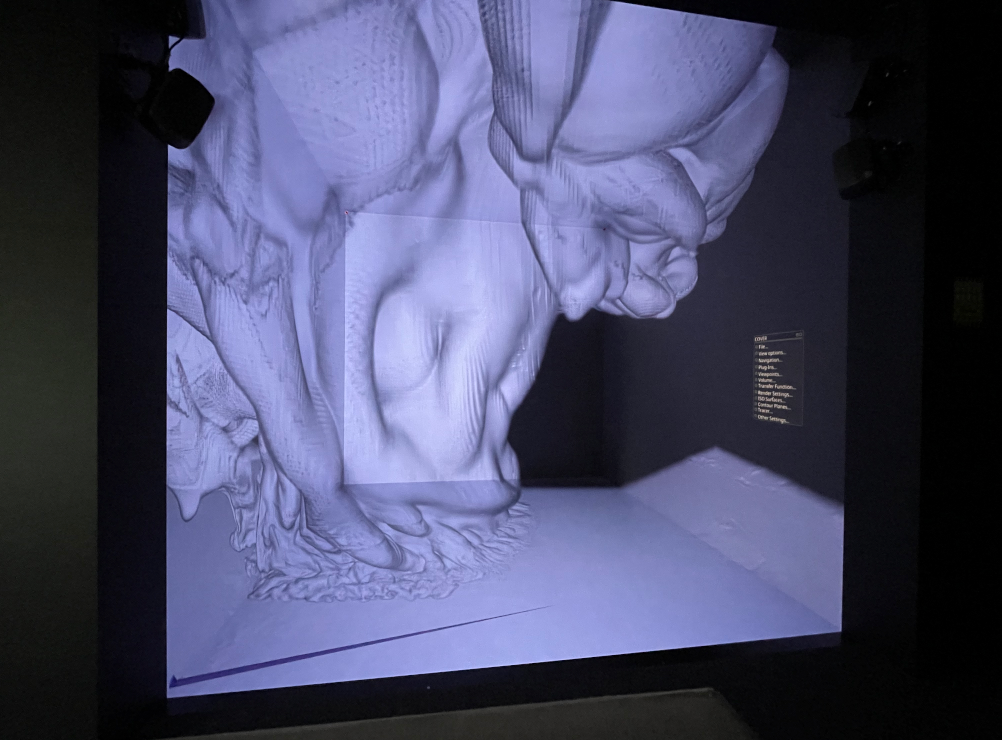}
            \caption[]%
            {{\small LANL Meteor Impact viewpoint~2 (zoomed-in)}}    
            \label{fig:LANL_meteor_impact_viewpoint_2}
        \end{subfigure}
        \caption[ The average and standard deviation of critical parameters ]
        {\small Photos taken in the CAVE at the University of Cologne, illustrating the data sets and viewports chosen for the evaluation.}

        \label{fig:photos_vp}
    \end{figure*}

\begin{figure}[htpb]
    \centering
    \begin{subfigure}[b]{0.475\textwidth}
        \centering
        \includegraphics[width=\textwidth]{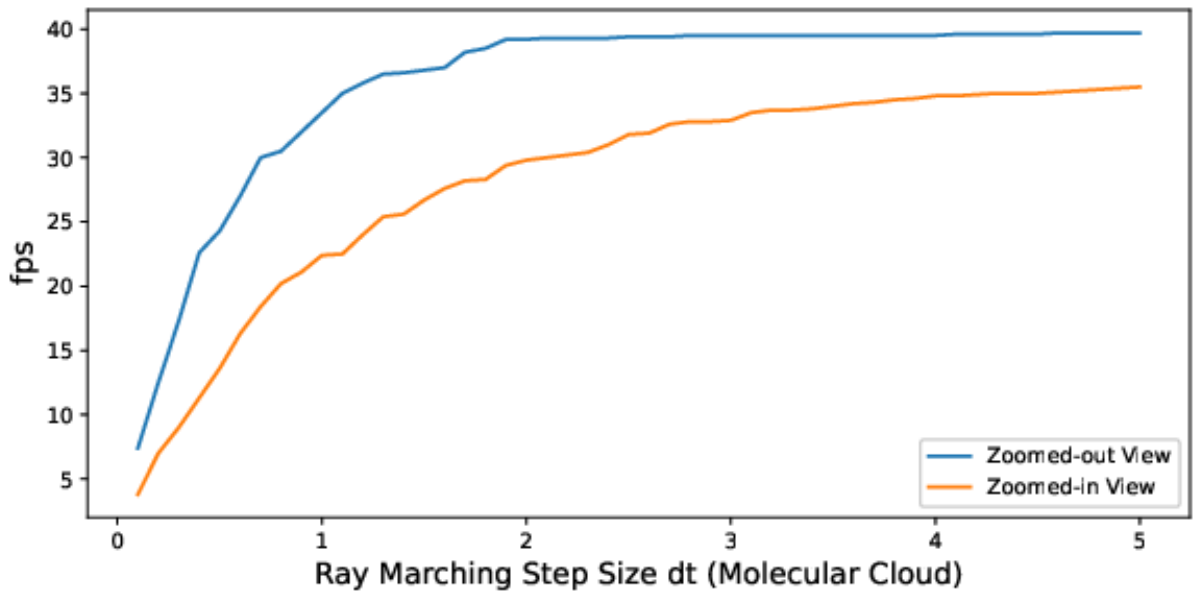}
        \caption{Molecular Cloud fps vs.\ dt.}
        \label{fig:molecular_cloud_result}
    \end{subfigure}
    \vspace{-0.5\baselineskip} 
    \begin{subfigure}[b]{0.475\textwidth}
        \centering
        \includegraphics[width=\textwidth]{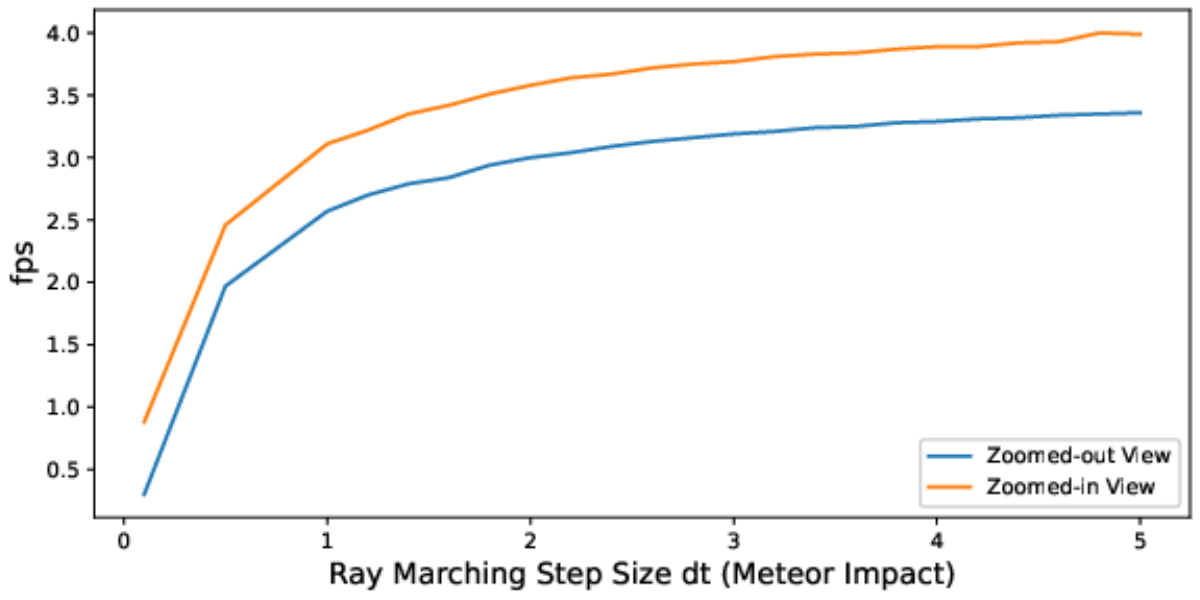}
        \caption{LANL Meteor Impact (at t=46,112s) fps vs.\ dt.}
        \label{fig:lanl_meteor_result}
    \end{subfigure}
    \hfill
    \caption{Rendering performance in frames per second (fps) vs. sampling rate, controlled by the ray marching step size $dt$.}
    \label{fig:graph_result}
    \vspace{-1em}
\end{figure}
We have tested the performance of the plugin on two AMR data sets:
a) SILCC Molecular Cloud data set: as presented in~\cite{girichidis2016silcc}, converted to the cells and scalar format using the FLASH converter coming with ExaBrick~\cite{wald2021exabrick}.
b) LANL Meteor Impact: Meteor impact simulation data set (at time t=46,112s) produced by Gisler et al.\ at Los Alamos National Laboratory (LANL)~\cite{patchett2016visualization} using the xRage simulation code~\cite{gittings2008rage}. 
\begin{table}
\centering
\begin{tabular}{r|ccc|c}
\toprule
data set & \#~Cells & \#~Bricks & \#~ABRs & Size on Disc\\
\midrule
SILCC & 15.8~M & 40 & 191 & 363~MB \\
Impact & 283~M & 3.1~M & 77.1~M & 6.5~GB \\
\bottomrule
\end{tabular}
\caption{\label{tab:datasets}%
Statistics for the data sets we use for the evaluation.}
\vspace{-2em}
\end{table}
Statistics for these data sets are found in \cref{tab:datasets}.
We picked two reprentative viewports (one zoomed-in view, and a zoomed-out overview where the camera is moved outside the scene bounds) depicted in \cref{fig:photos_vp}.

We are interested in the sampling quality---represented by the ray marching
base step size $dt$---that can be achieved while retaining real-time framerates.
Since ExaBrick employs adaptive sampling, $dt$ serves as a scaling factor rather
than an absolute step size; in cases where we sample an ABR with the finest
resolution, e.g., $dt=1.0$ means we are marching through that region with steps of
size one; when sampling an ABR where the finest cells are from refinement level $L=1$,
$dt=1.0$ means the step size is $2^L$ $(=2)$, etc. Also note that in the CAVE setup,
the framerate will be limited by that of the slowest viewport, i.e., the framerates
we report, in frames per second (fps), are that of the most costly frame in the whole
setup of stereo views across the five CAVE
projection planes.

For the molecular cloud data set, a total of 50 measurements were
taken across the range of $dt=0.1$ to $dt=5.0$.
For the LANL meteor impact data set, 23 measurements were recorded spanning $dt=0.1$ to $dt=5.0$.
We report our results in \cref{fig:graph_result}.

The results suggest that the molecular cloud data set can be
rendered at interactive rates (25~fps and higher) when setting
base $dt=1$. This is encouraging, as on average, at this quality setting we will sample
each cell exactly once. We also observe that the viewpoint dependency
is neglible. We achieve relatively consistent framerates regardless which
zoom level we choose. We observe diminishing returns on performance
when choosing base $dt>2$, but note that the quality is impacted significantly
at such low sampling rates.

For the LANL meteor impact, we observe similar scalability curves when
varying base $dt$, with the same plateau at $dt=2$, yet we note that framerates
are not interactive enough for virtual reality in the CAVE. The maximum framerate we achieve is about 3-4 fps; we only achieve higher framerates (not included in our measurements) when zooming out significantly; this points to
future work on finding different trade-offs between quality and performance other
than simply varying $dt$, or incorporating more sophisticated latency hiding techniques. We conclude
that framerates depend largely on the size and complexity of the data set.

\section{Discussion} \label{sec:discussion}


The research presented in this paper makes several significant contributions to the field of real-time rendering of large AMR data in VR. 
Firstly, the developed ExaBrick plugin successfully enables the real-time visualization and exploration of complex AMR data sets, leveraging the VR visualization system of COVISE and interactive navigation techniques. The integration of ExaBrick as a COVISE plugin provides an efficient and scalable framework for rendering large-scale AMR data in VR environments. Additionally, the methods to adjust the ray marching step size guarantees a balanced trade-off between rendering speed and visual fidelity.
Overall, this work offers scientists and researchers a powerful tool for exploring, analyzing and understanding complex AMR data sets.

While the implemented solution demonstrates promising results, it is important to acknowledge its limitations. One limitation is the performance impact when dealing with very large AMR data sets like the LANL Meteor data set, where further tests and optimization may be required to maintain real-time rendering performance.
Our observations point to future work.

To overcome limitations with larger data sets, a sensible approach could be the use of level-of-detail techniques,
so the sampling rate is not only chosen based on the local AMR level, but also on
proximity to the viewing position. Other countermeasures might include latency hiding
techniques that generate novel views from outdated frames while waiting for new frames
to be rendered.

Future work may also involve improving the visual fidelity, though within the constraints that interactivity allows.
It would, e.g., be interesting to incorporate global illumination techniques as
in~\cite{zellmann-beyond-exabricks}; however, as these techniques are notoriously
expensive, would require even more aggressive optimization such as using Monte Carlo
denoising techniques.

We finally note that both the original ExaBrick library, COVISE and OpenCOVER, as well as our
plugin, are open source and published under the links provided above, allowing researchers
to use our software in their own virtual environments.

\section{Conclusion} \label{sec:conclusion}

In conclusion, the integration of the ExaBrick framework as a plugin within the COVISE visualization system provides a promising solution for real-time visualization of large AMR data sets in VR environments. The plugin demonstrates efficiency in rendering moderately sized data sets and offers interactive exploration experiences. While challenges with extremely large data sets remain, further optimizations and advancements in rendering techniques could enhance the plugin's capabilities. Overall, the ExaBrick plugin stands as an important tool for immersive scientific visualization, with the potential to evolve further as technology progresses.

\section*{Acknowledgments}
\ifsubmission
Intentionally omitted for review.
\else

This work was supported by the Deutsche Forschungsgemeinschaft (DFG, German
Research Foundation) under grant no.~456842964. We are grateful to the IT Center Cologne at the University of Cologne for generously granting us access to the CAVE VR environment. Our appreciation extends to the creators of the ExaBrick framework and COVISE for laying down robust foundations upon which this work was built.
\fi

\bibliographystyle{eg-alpha-doi}  
\bibliography{main}

\newcommand{\etalchar}[1]{$^{#1}$}
\begin{thebibliography}{\uppercase{SDWWL01}}

\bibitem[BC89]{berger:1989:local}
\textsc{Berger M.~J., Colella P.}:
\newblock {Local} {Adaptive} {Mesh} {Refinement} for {Shock} {Hydrodynamics}.
\newblock \emph{{Journal} of {Computational} {Physics} 82}, 1 (1989), 64--84.

\bibitem[BO84]{berger:1984:adaptive}
\textsc{Berger M.~J., Oliger J.}:
\newblock {Adaptive} {Mesh} {Refinement} for {Hyperbolic} {Partial}
  {Differential} {Equations}.
\newblock \emph{{Journal} of {Computational} {Physics} 53}, 3 (1984), 484--512.

\bibitem[BWG11]{burstedde:2011:p4est}
\textsc{Burstedde C., Wilcox L.~C., Ghattas O.}:
\newblock {p4est}: {Scalable} {Algorithms} for {Parallel} {Adaptive} {Mesh}
  {Refinement} on {Forests} of {Octrees}.
\newblock \emph{{SIAM} {Journal} on {Scientific} {Computing} 33}, 3 (2011),
  1103--1133.

\bibitem[CGL{\etalchar{*}}09]{colella:2009:chombo}
\textsc{Colella P., Graves D.~T., Ligocki T., Martin D., Modiano D., Serafini
  D., Van~Straalen B.}:
\newblock {Chombo} {Software} {Package} for {AMR} {Applications} {Design}
  {Document}.
\newblock \emph{{Available} at the {Chombo} website: http://seesar. lbl.
  gov/ANAG/chombo/(September 2008)} (2009).

\bibitem[CNSD{\etalchar{*}}92]{cruz-neira}
\textsc{Cruz-Neira C., Sandin D.~J., DeFanti T.~A., Kenyon R.~V., Hart J.~C.}:
\newblock {The} {CAVE}: {Audio} {Visual} {Experience} {Automatic} {Virtual}
  {Environment}.
\newblock \emph{{Commun.} {ACM} 35}, 6 (jun 1992), 64–72.
\newblock URL: \url{https://doi.org/10.1145/129888.129892}, \href
  {https://doi.org/10.1145/129888.129892} {\path{doi:10.1145/129888.129892}}.

\bibitem[GWC{\etalchar{*}}08]{gittings2008rage}
\textsc{Gittings M., Weaver R., Clover M., Betlach T., Byrne N., Coker R.,
  Dendy E., Hueckstaedt R., New K., Oakes W.~R., et~al.}:
\newblock {The} {RAGE} {Radiation-Hydrodynamic} {Code}.
\newblock \emph{{Computational} {Science} {\&} {Discovery} 1}, 1 (2008),
  015005.

\bibitem[GWN{\etalchar{*}}16]{girichidis2016silcc}
\textsc{Girichidis P., Walch S., Naab T., Gatto A., W{\"u}nsch R., Glover
  S.~C., Klessen R.~S., Clark P.~C., Peters T., Derigs D., et~al.}:
\newblock {The} {SILCC} ({SImulating} the {LifeCycle} of {molecular} {Clouds})
  project--{II}. {Dynamical} {evolution} of the {supernova-driven} {ISM} and
  the {launching} of {outflows}.
\newblock \emph{{Monthly} {Notices} of the {Royal} {Astronomical} {Society}
  456}, 4 (2016), 3432--3455.

\bibitem[HSM{\etalchar{*}}18]{horan:2018:cave-like}
\textsc{Horan B., Sevedmahmoudian M., Mortimer M., Thirunavukkarasu G.~S.,
  Smilevski S., Stojcevski A.}:
\newblock {Feeling} {Your} {Way} {Around} {a} {CAVE-Like} {Reconfigurable} {VR}
  {System}.
\newblock In \emph{2018 11th {International} {Conference} on {Human} {System}
  {Interaction} ({HSI})} (2018), pp.~21--27.
\newblock \href {https://doi.org/10.1109/HSI.2018.8431365}
  {\path{doi:10.1109/HSI.2018.8431365}}.

\bibitem[ICC23]{cave-cologne}
\textsc{{IT} {Center}~{Cologne} I.}:
\newblock {CAVE} {VR} {Environment} at the {University} of {Cologne}, 2023.
\newblock URL:
  \url{https://rrzk.uni-koeln.de/en/hpc-projects/visualization/cave}.

\bibitem[KHYD22]{koger2022virtual}
\textsc{Koger C.~R., Hassan S.~S., Yuan J., Ding Y.}:
\newblock {Virtual} {Reality} for {Interactive} {Medical} {Analysis}.
\newblock \emph{{Frontiers} in {Virtual} {Reality} 3} (2022), 782854.

\bibitem[KK19]{kalarat2019real}
\textsc{Kalarat K., Koomhin P.}:
\newblock {Real-time} {volume} {rendering} {interaction} in {Virtual}
  {Reality}.
\newblock \emph{{International} {Journal} of {Technology} 10}, 7 (2019),
  1307--1314.

\bibitem[KSH03]{kahler2003interactive}
\textsc{K{\"a}hler R., Simon M., Hege H.-C.}:
\newblock {Interactive} {Volume} {Rendering} of {Large} {Sparse} {Data} {Sets}
  {Using} {Adaptive} {Mesh} {Refinement} {Hierarchies}.
\newblock \emph{{IEEE} {Transactions} on {Visualization} and {Computer}
  {Graphics} 9}, 3 (2003), 341--351.

\bibitem[KWAH06]{kahler2006gpu}
\textsc{K{\"a}hler R., Wise J., Abel T., Hege H.-C.}:
\newblock {GPU}-assisted {Raycasting} for {Cosmological} {Adaptive} {Mesh}
  {Refinement} {Simulations}.
\newblock In \emph{{VG}@ {SIGGRAPH}} (2006), pp.~103--110.

\bibitem[MWUP20]{morrical:20:bilinear-elements}
\textsc{Morrical N., Wald I., Usher W., Pascucci V.}:
\newblock {Accelerating} {Unstructured} {Mesh} {Point} {Location} {With} {RT}
  {Cores}.
\newblock \emph{{IEEE} {Transactions} on {Visualization} and {Computer}
  {Graphics} PP} (12 2020), 1--1.
\newblock \href {https://doi.org/10.1109/TVCG.2020.3042930}
  {\path{doi:10.1109/TVCG.2020.3042930}}.

\bibitem[NASN23]{ospray-immersive}
\textsc{Nam J.~W., Abram G.~D., Samsel F., Navrátil P.~A.}:
\newblock {Immersive} {OSPRay}: {Enabling} {VR} {Experiences} with {OSPRay}.
\newblock In \emph{{PEARC}: {Practice} and {Experience} in {Advanced}
  {Research} {Computing} ({PEARC} 2023)} (2023).
\newblock preprint.
\newblock URL:
  \url{https://jungwhonam.github.io/images/publications/PEARC2023_Immersive-OSPRay.pdf},
  \href {https://doi.org/10.1145/3569951.3597579}
  {\path{doi:10.1145/3569951.3597579}}.

\bibitem[PBD{\etalchar{*}}10]{parker2010optix}
\textsc{Parker S.~G., Bigler J., Dietrich A., Friedrich H., Hoberock J., Luebke
  D., McAllister D., McGuire M., Morley K., Robison A., et~al.}:
\newblock {Optix}: {A} {General} {Purpose} {Ray} {Tracing} {Engine}.
\newblock \emph{{Acm} {transactions} on {graphics} ({tog}) 29}, 4 (2010),
  1--13.

\bibitem[PD09]{perret:09:inca6d}
\textsc{Perret J., Dominjon L.}:
\newblock {The} {INCA} {6D}: {A} {Commercial} {Stringed} {Haptic} {System}
  {Suitable} for {Industrial} {Applications}.

\bibitem[PST{\etalchar{*}}16]{patchett2016visualization}
\textsc{Patchett J., Samsel F., Tsai K.~C., Gisler G.~R., Rogers D.~H., Abram
  G.~D., Turton T.~L.}:
\newblock {Visualization} and {Analysis} of {Threats} from {Asteroid} {Ocean}
  {Impacts}.
\newblock \emph{{Los} {Alamos} {National} {Laboratory}} (2016).

\bibitem[RFL{\etalchar{*}}98]{rantzau1998covise}
\textsc{Rantzau D., Frank K., Lang U., Rainer D., W{\"o}ssner U.}:
\newblock {COVISE} in the {CUBE}: an {Environment} for {Analyzing} {Large} and
  {Complex} {Simulation} {Data}.
\newblock In \emph{{Proceedings} of the 2nd {Workshop} on {Immersive}
  {Projection} {Technology}} (1998), vol.~2001.

\bibitem[RLL{\etalchar{*}}96]{rantzau1996collaborative}
\textsc{Rantzau D., Lang U., Lang R., Nebel H., Wierse A., Ruehle R.}:
\newblock {Collaborative} and {Interactive} {Visualization} in a {Distributed}
  {High} {Performance} {Software} {Environment}.
\newblock In \emph{{High} {Performance} {Computing} for {Computer} {Graphics}
  and {Visualisation}: {Proceedings} of the {International} {Workshop} on
  {High} {Performance} {Computing} for {Computer} {Graphics} and
  {Visualisation}, {Swansea} 3--4 {July} 1995} (1996), Springer, pp.~207--216.

\bibitem[SC03]{sherman03}
\textsc{Sherman W.~R., Craig A.~B.}:
\newblock \emph{{Understanding} {Virtual} {Reality}: {Interface},
  {Application}, and {Design}}.
\newblock Wiley, 2003.

\bibitem[SCRL20]{sarton:20:gpu-vis}
\textsc{Sarton J., Courilleau N., Remion Y., Lucas L.}:
\newblock {Interactive} {Visualization} and {On-Demand} {Processing} of {Large}
  {Volume} {Data}: {A} {Fully} {GPU-Based} {Out-of-Core} {Approach}.
\newblock \emph{{IEEE} {Transactions} on {Visualization} and {Computer}
  {Graphics} 26}, 10 (2020), 3008--3021.
\newblock \href {https://doi.org/10.1109/TVCG.2019.2912752}
  {\path{doi:10.1109/TVCG.2019.2912752}}.

\bibitem[SDWWL01]{schulze2001volume}
\textsc{Schulze-D{\"o}bold J., W{\"o}ssner U., Walz S.~P., Lang U.}:
\newblock {Volume} {rendering} in a {virtual} {environment}.
\newblock In \emph{{Immersive} {Projection} {Technology} and {Virtual}
  {Environments} 2001: {Proceedings} of the {Eurographics} {Workshop} in
  {Stuttgart}, {Germany}, {May} 16--18, 2001} (2001), Springer, pp.~187--198.

\bibitem[SZD{\etalchar{*}}23]{sarton:2023:state}
\textsc{Sarton J., Zellmann S., Demirci S., G{\"u}d{\"u}kbay U.,
  Alexandre-Barff W., Lucas L., Dischler J.-M., Wesner S., Wald I.}:
\newblock {State-of-the-art} in {Large-Scale} {Volume} {Visualization} {Beyond}
  {Structured} {Data}.

\bibitem[WBUK17]{wald2017cpu}
\textsc{Wald I., Brownlee C., Usher W., Knoll A.}:
\newblock {CPU} {Volume} {Rendering} of {Adaptive} {Mesh} {Refinement} {Data}.
\newblock In \emph{{SIGGRAPH} {Asia} 2017 {Symposium} on {Visualization}}
  (2017), pp.~1--8.

\bibitem[WCM12]{weber2012efficient}
\textsc{Weber G.~H., Childs H., Meredith J.~S.}:
\newblock {Efficient} {Parallel} {Extraction} of {Crack-Free} {Isosurfaces}
  from {Adaptive} {Mesh} {Refinement} ({AMR}) {Data}.
\newblock In \emph{{IEEE} {Symposium} on {Large} {Data} {Analysis} and
  {Visualization} ({LDAV})} (2012), IEEE, pp.~31--38.

\bibitem[WMHL65]{woodcock1965techniques}
\textsc{Woodcock E., Murphy T., Hemmings P., Longworth S.}:
\newblock {Techniques} {used} in the {GEM} {Code} for {Monte} {Carlo}
  {Neutronics} {Calculations} in {Reactors} and {Other} {Systems} of {Complex}
  {Geometry}.
\newblock In \emph{{Proc.} {Conf.} {Applications} of {Computing} {Methods} to
  {Reactor} {Problems}} (1965), vol.~557, {Argonne} {National} {Laboratory}.

\bibitem[WML{\etalchar{*}}22]{wald2022owl}
\textsc{Wald I., Morrical N., Lacewell D., Pisha L., Amstutz J., Zellmann S.}:
\newblock {OWL}: {A} {Node} {Graph} {"Wrapper"} {Library} for {OptiX} 7, 2022.

\bibitem[WQ10]{osg}
\textsc{Wang R., Qian X.}:
\newblock \emph{{OpenSceneGraph} 3.0: {Beginner's} {Guide}}.
\newblock {Packt} {Publishing}, 2010.

\bibitem[WWW{\etalchar{*}}19]{wang:18:iso-amr}
\textsc{Wang F., Wald I., Wu Q., Usher W., Johnson C.~R.}:
\newblock {CPU} {Isosurface} {Ray} {Tracing} of {Adaptive} {Mesh} {Refinement}
  {Data}.
\newblock \emph{{IEEE} {Transactions} on {Visualization} and {Computer}
  {Graphics}} (2019).

\bibitem[WZU{\etalchar{*}}21]{wald2021exabrick}
\textsc{Wald I., Zellmann S., Usher W., Morrical N., Lang U., Pascucci V.}:
\newblock {Ray} {Tracing} {Structured} {AMR} {Data} {Using} {ExaBricks}.
\newblock \emph{{IEEE} {Transactions} on {Visualization} and {Computer}
  {Graphics} 27}, 2 (2021), 625–634.

\bibitem[Zel14]{zellmann:phd}
\textsc{Zellmann S.}:
\newblock \emph{{Interactive} {high-performance} {volume} {rendering}}.
\newblock PhD thesis, University of Cologne, Aug. 2014.

\bibitem[ZSM{\etalchar{*}}22]{zellmann:2022:cise}
\textsc{Zellmann S., Seifried D., Morrical N., Wald I., Usher W., Law-Smith J.,
  Walch-Gassner S., Hinkenjann A.}:
\newblock {Point} {Containment} {Queries} on {Ray} {Tracing} {Cores} for {AMR}
  {Flow} {Visualization}.
\newblock \emph{{Computing} in {Science} {Engineering}} (2022), 1--1.
\newblock \href {https://doi.org/10.1109/MCSE.2022.3153677}
  {\path{doi:10.1109/MCSE.2022.3153677}}.

\bibitem[ZWL17]{visionaray}
\textsc{Zellmann S., Wickeroth D., Lang U.}:
\newblock {Visionaray}: {A} {Cross-Platform} {Ray} {Tracing} {Template}
  {Library}.
\newblock In \emph{2017 {IEEE} 10th {Workshop} on {Software} {Engineering} and
  {Architectures} for {Realtime} {Interactive} {Systems} ({SEARIS})} (2017),
  pp.~1--8.
\newblock URL: \url{https://ieeexplore.ieee.org/document/9183547}, \href
  {https://doi.org/10.1109/SEARIS41720.2017.9183547}
  {\path{doi:10.1109/SEARIS41720.2017.9183547}}.

\bibitem[ZWS{\etalchar{*}}22a]{zellmann2022design}
\textsc{Zellmann S., Wald I., Sahistan A., Hellmann M., Usher W.}:
\newblock {Design} and {Evaluation} of a {GPU} {Streaming} {Framework} for
  {Visualizing} {Time-varying} {AMR} {Data}.
\newblock In \emph{{Eurographics} {Symposium} on {Parallel} {Graphics} and
  {Visualization}} (2022).

\bibitem[ZWS{\etalchar{*}}22b]{zellmann-beyond-exabricks}
\textsc{Zellmann S., Wu Q., Sahistan A., Ma K.-L., Wald I.}:
\newblock {Beyond} {ExaBricks}: {GPU} {Volume} {Path} {Tracing} of {AMR}
  {Data}, 2022.
\newblock \href {http://arxiv.org/abs/2211.09997} {\path{arXiv:2211.09997}}.

\end{thebibliography}

\balance

\end{document}